\documentstyle[pre,aps,psfig]{revtex}

\newcommand{\be}{\begin{equation}}
\newcommand{\ee}{\end{equation}}
\newcommand{\bea}{\begin{eqnarray}}
\newcommand{\eea}{\end{eqnarray}}
\newcommand{\es}{\ell_{\rm ES}}
\newcommand{\ld}{\ell_{\rm D}}
\newcommand{\jpa}{\vec j_s^{\|}}
\newcommand{\jpe}{\vec j_s^{\perp}}
\newcommand{\js}{\vec j_s}
\newcommand{\jt}{\vec j_t}
\newcommand{\mpe}{\vec m_\perp}
\newcommand{\jx}{\vec j_{[100]}}
\newcommand{\jxy}{\vec j_{[110]}}

\begin{document}

\title{Crystal symmetry, step-edge diffusion and unstable growth}

\author{Paolo Politi$^{1,2,}$\footnote{Corresponding author. 
Present address: INFM, L.go E.~Fermi 2, I-50125 Florence.
E-mail: {\tt politi@fi.infn.it}} and Joachim Krug$^1$}

\address{$^1$Fachbereich Physik, Universit\"at GH Essen,
45117 Essen, Germany}
\address{$^2$Istituto Nazionale per la Fisica della Materia,
Unit\`a di Firenze,\\ L.go E. Fermi 2, 50125 Firenze, Italy}

\date{\today}
\maketitle

\begin{abstract}
We study the effect of crystal symmetry and step-edge diffusion on the
surface current governing the evolution of a growing crystal surface.
We find there are two possible contributions to anisotropic currents,
which both lead to the destabilization of the flat surface:
terrace current $\vec j_t$, which is parallel to the slope $\vec m
=\nabla z(\vec x,t)$, and step current $\vec j_s$, which has
components parallel $(\jpa)$ and perpendicular $(\jpe)$ to the slope.
On a high-symmetry surface, terrace and step currents are generically 
{\em singular} at zero slope, and this does not allow to perform
the standard linear stability analysis. 
As far as a one-dimensional profile is considered,
$\jpe$ is irrelevant and $\jpa$ suggests that mound sides align along 
$[110]$ and $[1\bar 10]$ axes.
On a vicinal surface, $\js$ {\em destabilizes} against step bunching;
its effect against step meandering depends on the step orientation,
in agreement with the recent findings by O.~Pierre-Louis et al.
[Phys. Rev. Lett. {\bf 82}, 3661 (1999)].
\end{abstract}

\section{Introduction}

The kinetic stability of a crystal growing 
homoepitaxially by Molecular Beam Epitaxy is
determined primarily by the possible existence of a slope-dependent
mass current $\vec j(\vec m)$ along the surface, 
i.e. by a current which does not vanish in the limiting 
case of a constant slope $\vec m$ ($\vec m=\nabla z$, where $z(\vec r,t)$ 
is the local height) \cite{villain,kps,reviews}. 
Such a current is generally ascribed to 
the so-called Ehrlich-Schwoebel (ES) effect at step-edges, which 
hinders interlayer diffusion \cite{schwoebel}. 

On singular surfaces, experimental results (mainly on metal growth)
show that the instability leads to mound formation and often to a
coarsening process, where the typical size $L$ of the mounds increases in time
(generally with a power law: $L(t)\sim t^n$) \cite{reviews}.
The template of the mound structure is already formed in the
early stages of growth (the so-called `linear regime'), and here 
crystal structure should determine shape and orientation of mounds. 
For example, cubic crystals are characterized by a four-fold  and
a six-fold symmetry, respectively on (100)  and (111) faces:
experimental analysis by Scanning Tunneling Microscopy 
has indeed shown square based mounds on
Fe and Cu(100)~\cite{Stroscio93,Thurmer,Zuo} and triangular based ones on 
Rh and Pt(111)~\cite{Tsui,Kalff}.
The relevance of the in-plane symmetry for the later stages of the
growth process has been
definitely proven by Siegert~\cite{S98}, who has shown --through a
continuum description of the surface-- that unstable currents with
different in-plane symmetries may give rise to different coarsening 
exponents $n$.
For vicinal surfaces ES barriers at steps are known
to stabilize against step bunching and to destabilize against
step meandering \cite{BZ,pimp,OPL}.

It is therefore extremely important to determine what are the
microscopic mechanisms giving rise to slope-dependent currents $\vec j$,
what is the expression of $\vec j$, and how lattice symmetry enters in it.
One of the main results of the present paper is the finding of {\em two}
contributions to the slope-dependent current,  one due to terrace
diffusion ($\vec j_t$) and one due to step diffusion ($\vec j_s$).
Both contributions are  {\em singular} at zero slope\footnote{%
In the limit of very large ES effect, $\vec j_t$ is no more anisotropic
and therefore no more singular at $\vec m=0$.}.
This is at odds with the usual
phenomenological expressions of $\vec j_t$, used in the
continuum description of surface growth \cite{S98,siegert94,siegert97}, which all
reduce to the simple isotropic form
$\vec j_t = a \vec m$ in the small slope regime ($\vec m\to 0$).
We will see that our expressions for $\vec j_t$ and $\vec j_s$
remain anisotropic even in
this limit, and that implies a singular behaviour in $\vec m=0$. 
Other important results concern the step current $\vec j_s$, which
is found to destabilize layer-by-layer growth against mound formation on a
high symmetry surface, and step-flow against step bunching on a 
vicinal surface. 
Step-flow is stable or unstable against step meandering, 
depending on the step orientation.

The destabilizing effect of $\js$
on a singular surface has been observed
independently by O.~Pierre-Louis et al.~\cite{Einstein} and by Ramana Murty 
and Cooper~\cite{Ramana}. The former have also studied analytically the effect
on step meandering. Here we provide a unified treatment 
of these diverse effects within 
a continuum description of the surface, we predict the new phenomenon of 
step bunching induced by step currents, and we analyze the different 
anisotropic behaviours of $\js$ and $\jt$.

\section{Terrace current}

\begin{figure}
\centerline{\psfig{figure=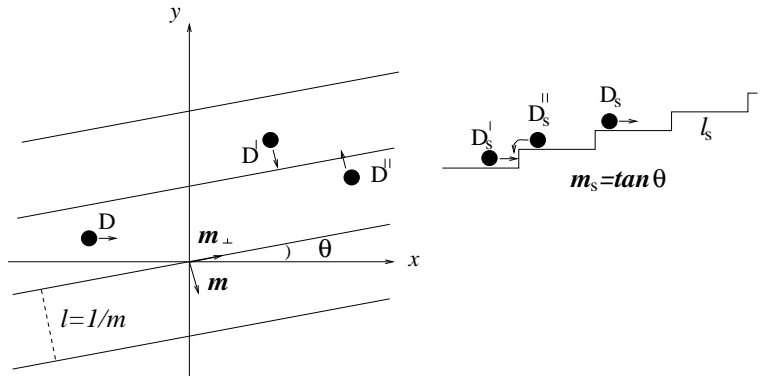,width=10cm,angle=-90}}
\caption{Sequence of equispaced steps, typical of a vicinal surface
(left) and just a single step (right), taking into account the discrete
character of the lattice. All the symbols are explained in the text.}
\label{sup-vic}
\end{figure}

In Fig.~\ref{sup-vic} we draw a piece of a vicinal surface corresponding
to a constant slope $\vec m$, and a piece of a step.
Once adatoms have landed on the surface, they perform
a diffusion process until they stick to the upper or lower
step. The attachement rate $D'$ from below is considered extremely fast 
($D'/D=\infty$, $D$ being the diffusion constant on the terrace); 
the rate $D''$ from above defines
the ES length $\es=(D/D'' - 1)$ (in units of the lattice spacing) 
\cite{reviews,pv96,Psolo}. 
This should be compared to the diffusion
length $\ld$ representing the minimal distance between nucleation
centers on a high-symmetry surface \cite{submono}.
Under the usual conditions of crystal growth we have $\ld\gg 1$, while both the
cases $\es\ll\ld$ (weak ES effect) and $\es\gg\ld$ (strong ES effect)
may take place \cite{reviews}.

In one dimension (1d) we write the ES current (due to terrace
diffusion) as $j_t=mf(m^2)$, 
and at small slopes (in the sense that $m \ll 1/\ld$) 
we have the linear behaviour $j_t=a m\equiv f(0)m$, with~\cite{Psolo} 
$a=F\es\ld^2/[2(\es+\ld)]$, $F$ being the intensity of the external flux
(i.e. the number of particles landing on the surface per unit time and lattice
site). In two dimensions (2d), if we neglect
in-plane anisotropy we can generalize and write $\vec j_t=\vec m f(m^2)$.
Let us now discuss the microscopic origin of  anisotropy and
how it modifies $\vec j_t$. Throughout we will consider a (100)-surface
with fourfold symmetry, and take the $x$ ($y$) axis along the 
[100] ([010]) orientation, denoting by $\hat x$ and $\hat y$ the
corresponding in-plane unit vectors. 
The extension to other crystal symmetries is 
straightforward in principle.

In the absence of surface reconstructions, terrace 
diffusion by itself is an isotropic process, at least in its
continuum description. In contrast, the sticking of an adatom to a
step depends on the microscopic environment, which 
depends on the step orientation.
So, the ES barrier seen by an adatom approaching a step depends on
the orientation of the surface and this dependence translates into an
orientation-dependent ES length $\es=\es(\theta)$, where
$\theta = \arctan (m_x/\vert m_y \vert)$ is the angle of the 
step relative
to the $x$-axis. Assuming straight steps, the 
expression for a one-dimensional surface can be taken over, and for
small slopes ($m\ld \ll 1$) we obtain
\be
\label{jt2d}
\vec j_t = a(\theta) \vec m = {F\es(\theta)\ld^2\over
2(\es(\theta)+\ld)} \vec m~.
\ee
The coefficient $a$ becomes independent of $\theta$ only in the 
regime of strong ES barriers, $\es(\theta)\gg\ld$
(in this limit, $a=F\ld^2/2$). For
weak barriers in-plane anisotropy is therefore present even in the `linear'
regime $m \ld \ll 1$ ($a=F\es(\theta)\ld/2$).
Through the dependence of $\theta$ on $\vec m$,
Eq.(\ref{jt2d}) is manifestly non-analytic at $\vec m=0$.

\section{Step current}
\label{Stepcurrent}

Next show that
crystal symmetry manifests itself also (and perhaps mainly) through
step diffusion.
Once adatoms have reached a step, they can diffuse along 
it at a rate $D_s$ and stick to a kink edge (see Fig.~\ref{sup-vic}).
Similarly to terrace diffusion, only if there is an
asymmetry between $D_s'$ and $D_s''$,
a net step current $\vec j_s$ exists; the strength of the
asymmetry determines an ES length along the step which 
will be called $\ell_k$, the subscript $k$ standing for kink. 
Step diffusion biased by kink barriers is 
similar to terrace diffusion along a one-dimensional surface \cite{Einstein},
but some differences are worth to be stressed. 

i)~All the possible 
in-plane orientations $\theta$ of the step, with the correct symmetries, 
should be taken into account,
because --especially for a high-symmetry orientation--
all the $\theta$ are found on the same surface.
This may be true also for a vicinal surface, if steps are subject to a
strong meandering~\cite{OPL}. In particular, orientations corresponding to
$\theta=0$ and $\theta=\pi/4$ will be seen to behave in a
qualitatively different way.  
ii)~Adatoms arrive at 
the step at a rate $F_s$ depending on the terrace size $\ell$.
For equally spaced steps $F_s=F\ell$. 
However, since (in 2d) the surface current is defined as the
number of atoms crossing per unit time a segment of unit length, orthogonal
to the current, the actual expression for the current is obtained
by multiplying the `single-step current' by the number of steps per
unit length, i.e. $1/\ell$. This factor cancels the factor $\ell$ 
appearing in $F_s$, since the current is proportional to $F_s$ as well.
iii)~A step is a one-dimensional object, and therefore
it has a larger roughness than a two-dimensional surface.
In the expression for the unstable current, the diffusion length gives
the minimal distance between steps (in 2d) or kinks (in 1d) along a
high-symmetry orientation. In 2d, steps are created by nucleation and
growth, and $\ld$ is generally given by an expression as
$\ld\approx (D/F)^\gamma$ with the exponent $\gamma$ depending on 
the details of the nucleation process \cite{submono}. 
In 1d, the corresponding
expression $\ell_{D_s} = (D_s/F_s)^{\gamma_s}$ should be compared to
the distance between thermally excited kinks, and the smaller one 
(called $\ell_d$) be chosen. In most of our discussion we will assume
that $\ell_d$ is sufficiently large so that double or multiple kinks
can be neglected. 

The high symmetry in-plane orientations [100] and [110] for a step are
fairly different in the mechanisms giving rise to a step-edge current.
Along a [100] segment, step diffusion takes place between nearest
neighbours lattice sites at a rate $D_s$, and the analogy with a one
dimensional surface is appropriate. In particular, 
an asymmetry in the sticking rates
to a kink determines a net current along the straight segments of
the step, i.e. in the $\hat x$ direction; when $\theta\ne 0$ this current 
does not vanish and it has a component along
the slope $\vec m$, which will be seen to destabilize the flat surface.

\begin{figure}
\centerline{\psfig{figure=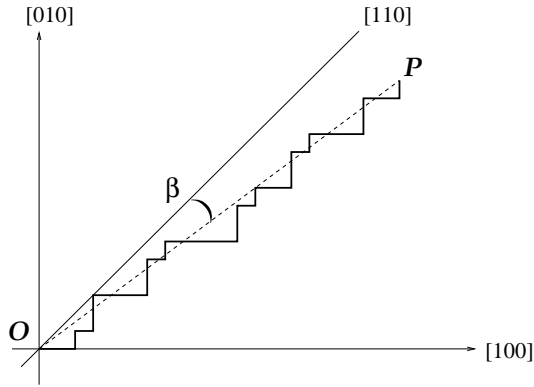,width=7cm,angle=-90}}
\caption{Step profile, when its average orientation is close to [110].}
\label{step}
\end{figure}

Conversely, along a [110] orientation, diffusion is a two-steps process 
and it is very much slower, because it requires detachment from a 
high coordination site. As a first approximation, it may even be
reasonable to assume that no detachment at all takes place.
This does not prevent a nonzero step-edge current, for the
following reason. Along the [100] orientation, kinks are due to
nucleation, or they must be thermally activated, because a kink increases
the total length of the step. Along and close to the [110] orientation,
the total length of the step does not depend on the specific sequence of
[100] and [010] terraces (see Fig.~\ref{step}), and therefore the
step is rough even at zero temperature. 
The path for going from the origin $O$ to $P$ is equivalent to a directed
random walk, where the asymmetry $p$ between the probabilities
$p_-=(1-p)/2$ and $p_+=(1+p)/2$ to go respectively in the $\hat x$ and
$\hat y$ directions, is nothing but the tangent of the angle 
$\beta = \pi/4 - \theta$ formed 
by the average orientation of the step with the [110] direction.
Since step diffusion does take place along [100] and [010] segments, each
step segment longer than one lattice constant contributes to the
$\hat x$ and $\hat y$ components of the step current.
This implies that $\vec j_s$ is nonzero also for $\theta=\pi/4$:
in this case $\vec j_s$ is exactly parallel to $\vec m$ and
it has a destabilizing character.
In the following we will consider separately the cases of small $\theta$
and $\theta$ close to $\pi/4$, and afterwards we will write down
a general expression for $\vec j_s$, valid for any value of the angle $\theta$.

For the moment, we will suppose that the slope $m=|\vec m|$ of the surface
is larger than $1/\ld$, i.e. we are in the `vicinal' regime.
The step current $\vec j_s$ can always be written as the sum
of $\jpa$ and $\jpe$, where $\jpa=(\js\cdot\vec m)\vec m/m^2$
and $\jpe=(\js\cdot\mpe)\mpe/m^2$. The vector $\mpe=(-m_y,m_x)$ is
orthogonal to $\vec m$.

If we are close to the [100] orientation, the current $\js$ is simply 
$\jx=j_{1d}(m_s)\hat x$, where $j_{1d}(m_s)$ is the usual unstable
current for a one-dimensional surface whose slope is $m_s=\tan\theta=
m_x/|m_y|$. For small $\theta$, $j_{1d}=a_s m_s$, with
$a_s=(F_s/\ell)\ell_k\ell_d^2/[2(\ell_k+\ell_d)]$. By decomposing $\jx$ along
$\vec m$ and $\mpe$, we obtain
\be
\jx = {j_{1d}(m_s)\over m^2}[m_x\vec m -m_y\mpe ].
\label{j100}
\ee
The uphill component is  $(\jx \cdot \vec m/m) \approx a_s m_x^2/m_y^2 > 0$,
for small $m_s$.
The positive value of this component explains why step-edge current is
enough to destabilize a flat, high symmetry surface. More details are
given in Sec.~\ref{sss}.

If we are close to the [110] orientation, as explained above, the current
originates from entropic fluctuations around the straight step,
which create segments along [100] and [010] directions. Each segment
of length $s$ contributes a local current proportional to $(s-1)$.
Therefore, the components $j_x$ and $j_y$ of the step current along the
$\hat x$ and $\hat y$ axes are simply proportional to the probabilities
(per step site) $p_-^2$ and $p_+^2$ to have a couple of consecutive
step sites in the horizontal and vertical direction, respectively:
\be
\jxy = {F_s\over\ell}(p_-^2\hat x - p_+^2\hat y) =
F\left( {1-p\over 2}\right)^2\hat x
- F\left( {1+p\over 2}\right)^2\hat y~~.
\ee

By using the relations $p=\tan(\theta -\pi/4)$ and $m_s=\tan\theta$, after
some easy algebra we obtain
\be
\jxy = {F\over\sqrt{1+m_s^2}}\left[
{|m_s|\over 1+|m_s|} {\vec m\over m} +
{1-|m_s|^3\over(1+|m_s|)^2} {\mpe\over m} \right]~.
\label{j110}
\ee

\begin{figure}
\centerline{\psfig{figure=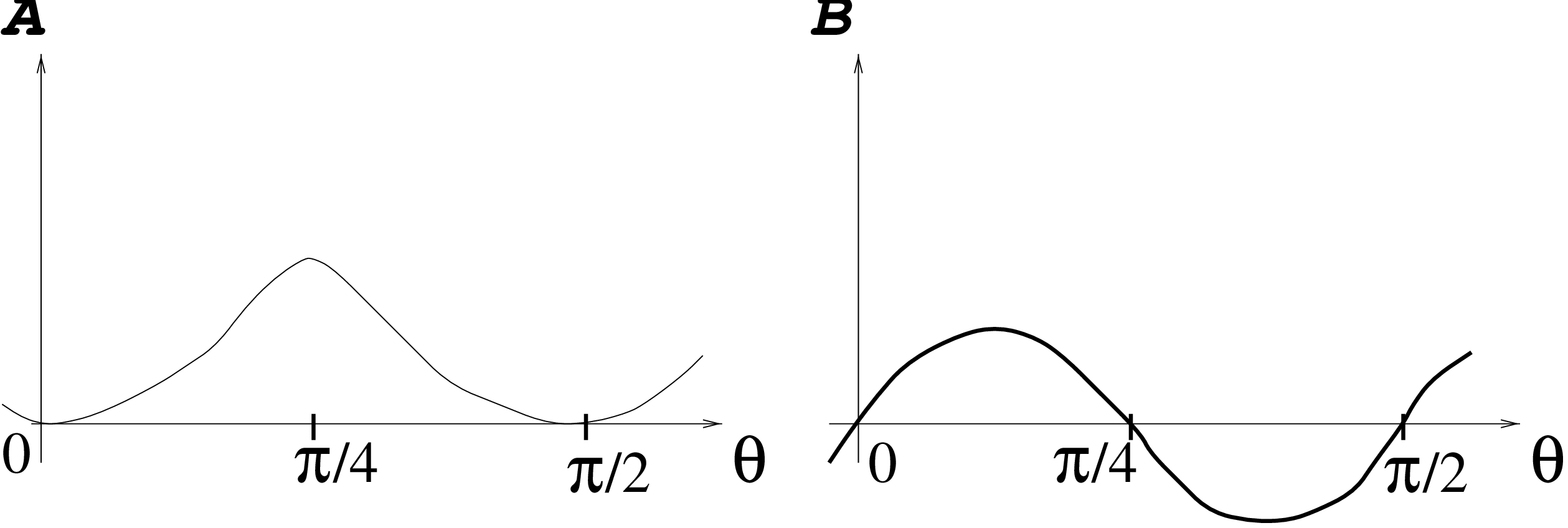,width=8cm,angle=-90}}
\caption{Plots of $A(\theta)$ and $B(\theta)$, which are defined in
Eq.~(\protect\ref{jsgeneral}).}
\label{figAB}
\end{figure}

The expressions (\ref{j100}) and (\ref{j110}) are valid close to the [100] 
and [110] orientations: they can be generalized to any
value of $m_s$, i.e. to any angle $\theta$, by writing
\be
\js = A(\theta)\vec m/m + B(\theta)\mpe/m~.
\label{jsgeneral}
\ee
Both $A$ and $B$ are periodic in $\theta$, with period $\pi/2$ 
(see Fig.~\ref{figAB}). Their
behaviours close to $\theta=0$ and $\theta=\pi/4$ are derivable from
$\jx$ (Eq.~(\ref{j100})) and $\jxy$ (Eq.~(\ref{j110})). 
More precisely, $A(\theta)$ is always nonnegative, 
it has a minumum for $\theta=0$ and a
maximum for $\theta=\pi/4$. The function $B(\theta)$ vanishes at $\theta=
0,\pi/4$, it has a positive slope in $\theta=0$ and a negative slope
in $\theta=\pi/4$. These properties are all that we need in the
following, and they do not depend on the specific model assumed
to derive Eqs.~(\ref{j100},\ref{j110}), because they are mainly due to
symmetry considerations. 
For example, if multiple kinks are allowed when the step is
close to the [100] orientation, $A(\theta)$ is no more zero at $\theta=0$, 
but $A(0)>0$ and $\theta=0$ is still a minimum.
Finally note that in contrast to the terrace current $\vec j_t$ \cite{reviews},
the step contribution (\ref{jsgeneral}) is independent\footnote{%
Strictly speaking,
a slope-dependence is maintained through the $\ell$-dependence
in $\ell_{D_s}$, but when $\ell$ is small, $\ell_{D_s}$ should be replaced by
the distance between thermally activated kinks (see the main text).}
of the surface slope $m$, i.e. the step distance, in the vicinal regime
($m\ld\gg 1$).

Before going on, let us generalize the expression of $\js$ to any value of the
surface slope $\vec m$. In the limit $m\ll 1/\ld$, both $\js$ and $\vec j_t$ 
go to zero, because contributions coming from steps and terraces 
of opposite sign compensate~\cite{reviews,siegert94}.
In a region of small slope, $\left.\vec j_{s,t}\right|_{\rm small~slope}=
{N_+-N_-\over N}\left.\vec j_{s,t}\right|_{\rm large~slope}$, where
$N_\pm$ is the number of positive and negative steps (for $\js$) or
terraces (for $\jt$) and $N=N_++N_-$. Since $(N_+-N_-)/N=m\ld$,
in the small slope regime: 
$\jpa = A(\theta)\ld\vec m$ and $\jpe = B(\theta)\ld\mpe$.
For a generic slope, we can write
\be
\jpa = A(\theta)g(m^2)\vec m ~~~~~~~~~~~~~
\jpe = B(\theta)g(m^2)\mpe
\label{jpp}
\ee
where $g(m^2)\to\ld$ for  $m\ld\ll 1$ and $g(m^2)\to 1/m$ for $m\ld\gg 1$.
The simplest function interpolating between the two limiting values is 
$g(m^2)=\ld/\sqrt{1+m^2\ld^2}$, 
but its actual form does not need to be specified.

\section{Stability of vicinal surfaces}

Let us now perform a linear stability analysis of a growing
vicinal surface of average slope $\vec m_0$. 
The local height is $z(\vec x,t)=\vec m_0\cdot\vec x
+\epsilon(\vec x,t)$ and the local slope is
$\vec m=\vec m_0 + \nabla\epsilon$. The evolution, {\em as determined by the
step current}, is given by the equation $\partial_t z = -\nabla\cdot\js$.
By using the general properties given above, for $A$ and $B$, and
working in the special cases $\theta_0=0$ and $\theta_0=\pi/4$, we obtain:
\be
\partial_t z = -\nabla\cdot\js =
-g(m_0^2)[A(\theta_0) + B'(\theta_0)]\partial^2_\perp \epsilon
-A(\theta_0)(\partial/\partial m_0)[m_0 g(m_0^2)]
\partial^2_\| \epsilon
\ee
where $\partial_\perp$ and $\partial_\|$ are directional derivatives
perpendicular and parallel to $\vec m$ (i.e. parallel and perpendicular to 
the step). Thus the coefficient of
$\partial^2_\perp$ gives informations on step meandering, and 
that of $\partial^2_\|$ on step bunching \cite{rsk96}. 

Since $A(\theta)\ge 0$ and $(\partial/\partial m_0)[m_0 g(m_0^2)]>0$,
the current $\js$  (more precisely $\jpa$) has always a destabilizing 
character against step bunching; if multikinks are not allowed along the
[100] orientation, $A(0)=0$ and the effect is absent along the
$\hat x,\hat y$ axes. 
Also, as it may be expected, $\jpe$ has no effect on this instability.

Concerning step meandering, we must distinguish between $\theta_0=0$
and $\theta_0=\pi/4$, because $B'(\theta_0)$ has opposite signs in the
two cases. For $\theta_0=0$ the derivative is positive and therefore
steps along the [100] orientation are destabilized by step meandering.
On the contrary, the evaluation of $[A(\pi/4) + B'(\pi/4)]$ gives,
using Eq.~(\ref{j110}), a negative result $(=-F/\sqrt{2})$.
Therefore, steps along the [110] orientations are stabilized against
step meandering. 

Our conclusion regarding the [100] steps agrees with the analysis
of Pierre-Louis et al.\cite{Einstein}, while in the case of [110] steps
they find stability (instability) for large (small) values of the
kink ES length $\ell_k$. Our expression (\ref{j110}) is valid if adatoms 
are not allowed to turn around corners and this effectively sets
$\ell_k = \infty$. 
Its generalization to any $\ell_k$ indeed shows that $A(\pi/4)$
becomes a minimum (and the quantity $[A(\pi/4) + B'(\pi/4)]$
changes sign) for $\ell_k<2\ell^*$, where $\ell^*$ is the typical distance
between corners along a [110] step. In the `random walk' model for the
step, $\ell^*=2$ and therefore the orientation [110] is stabilized unless
$\ell_k$ is extremely weak.

Ramana Murty and Cooper \cite{Ramana} have performed 
Monte~Carlo simulations of a vicinal surface, with 
steps along the [100] axis. Step meandering is indeed observed, even 
if the terrace current $\jt$ is absent. Conversely, no step bunching
seems to occur, suggesting that their simulations correspond to $A(0)=0$.

\section{Stability of singular surfaces}
\label{sss}

The analysis of a high-symmetry surface is complicated by the
non-analytic behavior of $\js$ and $\jt$ in $\vec m=0$.
In the small slope regime $(m\ll 1/\ld)$, $\jpa$ and $\jpe$
become (see Eq.~(\ref{jpp}))
\be
\jpa = \ld A(\theta)\vec m ~~~~~~~~
\jpe = \ld B(\theta) \mpe~.
\ee
It should be stressed that the singularity is physically justified, as we
now try to argue.
Close to an extremum of the profile, $z(x,y)=z_0 +(c_1/2)x^2 +
(c_2/2)y^2 + c_3 xy$, $\vec m=(c_1 x +c_3 y,c_2 y +c_3 x)$ 
and $m_s=(c_1 r+c_3)/|c_2 +c_3r|$, where $r=x/y$. The prefactors $A$ and
$B$ are therefore manifestly non-analytic functions at $x=y=0$.
The reason is that close to an extremum, steps are closed lines and
as the top (or the bottom) of the profile is approached, the step
orientation is no more defined. The angular dependence of $A$ and $B$ also
implies that the evolution equation does not become linear 
in the small-slope regime, and hence arbitrary profiles cannot
be treated as superpositions of harmonic ones. 

The problem of non-analyticity does not appear when we consider a
one-dimensional profile, i.e. a profile varying only in one direction
(for example, $z=z(x,t)$), because the prefactors $A$ and $B$
are now constants\footnote{%
The angle $\theta$ may indeed take the values $\theta_0$ and $(\theta_0 +\pi)$,
corresponding to $\tan\theta=\pm |m_s|$, but because of the $\pi/2$
periodicity, $A(\theta_0)=A(\theta_0 +\pi)$. This is no more true for the
(111) surface of a cubic crystal.}.
This implies that the divergence of the current is easily evaluated:
\bea
\partial_t z &=& -\nabla\cdot (\js + \jt) =
-\nabla\cdot [\ld A(\theta)\vec m + \ld B(\theta)\mpe + a(\theta)\vec m]
\nonumber \\
&=& -[\ld A(\theta_0) + a(\theta_0)]\nabla^2 z 
\equiv -\nu(\theta_0) \nabla^2 z ~.
\eea
The component of the step current parallel to the step $(\jpe)$ does not
contribute, because $\nabla\cdot\mpe\equiv 0$, while the component
parallel to the slope $(\jpa)$ destabilizes the flat surface,
analogously to $\jt$.

The instability gives rise to pyramid-shape mounds, whose orientations
$\theta_i$ should be the most unstable ones, i.e. correspond to the maxima of
$\nu(\theta)$.
In this respect, the step current favours the orientations forming
$45^\circ$ with the crystallographic axes, while the anisotropy induced
by the terrace current depends on the microscopic details of the
sticking processes.
Since the presence of more kinks along the step should favour the
descent of the adatom, it is reasonable to think that $\es(\theta)$ is
maximum in $\theta=0,\pi/2$ \cite{kinks},
and therefore the two contributions to $\nu(\theta)$ should compete. 
Close to $\theta=0$, we have
\be
\nu(\theta) = a_s m_s {m_x\over m} + a(\theta) = a_s\theta^2 + a(\theta)~.
\ee
By using expression (\ref{jt2d}) for $a(\theta)$ and the expression of
$a_s$ given above Eq.~(\ref{j100}), we obtain
\be
\nu''(0)=F\left[{\ld^3 \es''(0) \over 2(\es +\ld)^2} +
{\ell_k\ell_d^2 \over \ell_k +\ell_d}\right]~~.
\label{nu}
\ee

Mounds will align along the crystallographic axes if $\nu(\theta)$ has a 
maximum in $\theta=0$, i.e. if $\nu''(0)<0$.
It is apparent that for a sufficiently large ES effect at steps this
condition is not fulfilled, because
the anisotropic character of $\jt$ disappears. On the other hand,
$\es$ should also not be too small, otherwise $\jt$ itself would be
negligible.
Therefore $\nu(\theta)$ will have maxima in $\theta=0,\pi/2$ only if
several conditions are simultaneously satisfied:
i)~$\es''(0)$ must be negative\footnote{\label{foot}%
In simulations of solid-on-solid models the
step edge barriers are often implemented such as to reduce the
probability for adatoms to {\em approach} steps, rather than to
descend from them \cite{rsk96,pavel95}. In this case the
barrier at a [110] step may in fact (slightly) {\em exceed} that of
the close packed [100] step~\protect\cite{JKu}.};
ii)~$\es/\ld$ should be neither too large nor too small;
iii)~$\ell_k/\ell_d$ should be small.

Refs. \cite{Ramana} and \cite{Biehl} recently reported
simulations on high-symmetry surfaces, taking into account
step-edge diffusion. In both cases, if kink barriers are present the
mound sides align along [110] and equivalent axes. Since in ~\cite{Ramana}
there are no ES barriers at steps and in ~\cite{Biehl} the barriers are
infinite and therefore isotropic, mound orientation is determined only
by $\js$, in agreement with our picture.

\section{Discussion}

Some aspects of the subject considered in the present paper have not been
sufficiently clarified and they deserve further analysis.
First of all, the non-analytic behaviour of the surface current
remains problematic, because it implies that the continuum evolution
equation $\partial_t z + \nabla \cdot \vec j = 0$ is not defined
at $\vec m = 0$. As we have argued above, this non-analyticity is an
inescapable consequence of the persistence of crystal anisotropy in
the `linear' regime of the instability; if it could be removed,
e.g. through a more careful treatment of the interpolation between vicinal
and singular surfaces, this would also imply that mounds are initially
isotropic and develop their anisotropic shapes only in 
the nonlinear regime. 
It is however also conceivable that,
as in the case of equilibrium surface relaxation below the roughening 
temperature \cite{TR}, the appearance of a singularity
in the continuum evolution equation carries a real physical message:
That the surface is {\em not} well described by a smooth function 
$z({\vec r},t)$ near its maxima and minima. 

A second important aspect concerns vicinal surfaces.
We have seen that for [110] steps 
$\js$ has a stabilizing character with respect to
step meandering and a 
destabilizing character with respect to step bunching. It would be
interesting to evaluate quantitatively these effects and to compare
them with the opposing effects of the terrace current.
This comparison has been done for step meandering~\cite{Einstein},
and it seems that the effect of $\js$ may dominate $\jt$.
At any rate the predicted step bunching instability should be 
clearly visible in simulations of models which have 
no ES barriers but only asymmetric sticking at kinks \cite{Ramana}.

Finally, in this work we have not addressed the effects of crystal
anisotropy on the smoothening terms in the continuum evolution
equation, which are crucial in determining the actual length scale
of the instability \cite{reviews,rsk96}. Under far from equilibrium conditions,
the dominant smoothening mechanism is believed to be due to 
random nucleation \cite{pv96}, which is manifestly isotropic;
the anisotropy of the equilibrium step free energies will 
however be felt if detachment from steps becomes significant.   

\section{Conclusions}

In this paper we have studied the different contributions to the 
surface current on a (100)-surface, which depend only on the
slope $\vec m$. Such contributions come
from biased surface diffusion, both on terraces ($\jt$) and along steps
($\js$), where the bias mechanism is an Ehrlich-Schwoebel barrier at steps
and kinks, respectively.

The expressions of $\vec j_{s,t}$ are relevant in two respects: they
determine the linear stability of the flat surface and -- in the case
of unstable growth -- they also determine the shape and the orientation
of the emerging structure.

The terrace current is parallel to the slope, while the step current has
components parallel and perpendicular to the slope, because step
diffusion takes place along the [100]- and [010]-segments
that constitute the step.

A first important result is that the anisotropic character of $\js$ and
$\jt$ persists in the small slope regime $\vec m\to 0$:
this means that
they are both non-analytic at $\vec m=0$ and consequently this 
does not allow a complete description of the evolution of a high-symmetry 
surface.

Concerning the stability of a singular surface, $\js$ is found to be
destabilizing because it has a positive component in the direction of
$\vec m$. The most unstable orientations form angles of
45$^\circ$ with the
crystallographic axes, while $\jt$ is usually (but not always, see
footnote~\ref{foot}) thought to select the $\hat x,\hat y$ axes. 
If a competition exists $\js$ should prevail (see discussion below
Eq.~(\ref{nu})).

The stability of a vicinal surface is more complex.
The terrace current $\jt$ is known to
stabilize against step bunching and to destabilize towards
step meandering, whatever is the orientation of the surface.
Surprisingly, the step current is instead found to generally
favor step bunching
(along the crystallographic axes $\js$ has no
effect if multiple kinks are not present).
Finally the effect of $\js$ on step meandering strongly depends
on the surface orientation and the strength of the ES effect at kinks:
[100] steps are destabilized, while [110] steps may be stabilized if 
the kink ES-barrier is not too weak.

\vspace{0.5cm}

\noindent
{\bf Acknowledgements.} We have benefited from discussions with
Th. Michely and P. \v{S}milauer. P.P. acknowledges support by 
the Alexander von Humboldt foundation. The work of J.K. was supported
by DFG within SFB 237. 

\appendix

\section{List of symbols}

\begin{tabular}{l|l}
$z(\vec x,t)$ & local heigth of the surface at point $\vec x$ and time $t$ \\
$\vec m=\nabla z$ & local slope \\
$\mpe = (-m_y,m_x)$ & vector orthogonal to the slope \\
$\theta=\arctan\left({m_x\over |m_y|}\right)$ & 
angle between the average orientation of a step and the $\hat x$ axis \\
$m=|\vec m|=|\mpe|$ & modulus of the local slope \\
$\ell=1/m$ & terrace size \\
$\vec m_0$ & average slope of a vicinal surface \\
$\jt$ & terrace current \\
$\js$ & step current \\
$\jpa$ & step current parallel to the slope $\vec m$ \\
$\jpe$ & step current perpendicular to the slope $\vec m$ \\
$F$ & intensity of the external flux of particles \\
$F_s=F\ell$ & rate of particles arriving to the step \\
$D$ & diffusion constant on the terrace \\
$D_s$ & diffusion constant along a step \\
$D'$ & attachement rate of an adatom on a terrace to the ascending step \\
$D''$ & attachement rate of an adatom on a terrace to the descending step \\
$D_s'$ & attachement rate of an adatom on a step to the ascending kink \\
$D_s''$ & attachement rate of an adatom on a step to the descending kink \\
$\es$ & ES length on a terrace \\
$\ell_k$ & ES length along a step \\
$\ld$ & diffusion length on a terrace \\
$\ell_{D_s}$ & diffusion length along a step \\
$\ell_d$ & the minimum between $\ell_{D_s}$ and the distance between thermally\\
 & activated kinks \\
\end{tabular}

\begin{tabular}{l|l}
$p_\mp$ & In the directed random walk picture of a step close to the [110]
orientation, \\
& probabilities that the step goes in the $\hat x$ ($p_-$)
and $\hat y$ ($p_+$) directions \\
$p=p_+-p_-$ & asymmetry in the probabilities $p_\pm$ \\

\end{tabular}

\end{document}